\documentclass[sigconf]{acmart}
\pdfoutput=1

\acmConference[EASE 2025]{The 29th International Conference on Evaluation and Assessment in Software Engineering}{17–20 June, 2025}{Istanbul, Türkiye}
\graphicspath{{Images/}}
\usepackage{pifont}
\usepackage{amsmath,amssymb,amsfonts}
\usepackage{algorithmic}
\usepackage{graphicx}
\usepackage{textcomp}
\usepackage{multirow} 
\usepackage{xcolor}
\usepackage{hyperref}
\usepackage[flushleft]{threeparttable}
\usepackage{titlesec}

\begin{document}

\title{Reflection on Code Contributor Demographics and Collaboration Patterns in the Rust Community}

\author{Rohit Dandamudi}
\affiliation{%
  \institution{University of British Columbia}
  \city{Kelowna}
  \state{BC}
  \country{Canada}
}
\email{rohitd@student.ubc.ca}

\author{Ifeoma Adaji}
\affiliation{%
  \institution{University of British Columbia}
  \city{Kelowna}
  \state{BC}
  \country{Canada}
}
\email{ifeoma.adaji@ubc.ca}

\author{Gema Rodríguez-Pérez}
\affiliation{%
  \institution{University of British Columbia}
  \city{Kelowna}
  \state{BC}
  \country{Canada}
}
\email{gema.rodriguezperez@ubc.ca}

\begin{abstract}
Open source software communities thrive on global collaboration and contributions from diverse participants. This study explores the Rust programming language ecosystem to understand its contributors' demographic composition and interaction patterns. Our objective is to investigate the phenomenon of participation inequality in key Rust projects and the presence of diversity among them. We studied GitHub pull request data from the year leading up to the release of the latest completed Rust community annual survey in 2023. Specifically, we extracted information from three leading repositories: Rust, Rust Analyzer, and Cargo, and used social network graphs to visualize the interactions and identify central contributors and subcommunities. Social network analysis has shown concerning disparities in gender and geographic representation among contributors who play pivotal roles in collaboration networks and the presence of varying diversity levels in the subcommunities formed. These results suggest that while the Rust community is globally active, the contributor base does not fully reflect the diversity of the wider user community. We conclude that there is a need for more inclusive practices to encourage broader participation and ensure that the contributor base aligns more closely with the diverse global community that utilizes Rust.
\end{abstract}



\settopmatter{printfolios=true}
\maketitle

\section{Introduction}
\label{intro}
Open source software (OSS) communities form the backbone of contemporary software development, enabling global collaboration and innovation. Participants from diverse backgrounds contribute in various ways (i.e., reporting issues, submitting pull requests (PRs), conducting code reviews, and updating documentation). These participants often contribute in their free time and at a level that aligns with their expertise and availability, fostering an environment of collective growth and learning~\cite{lakhani2005hackers,motivation-in-oss}. Despite the seemingly vast pool of available developers, research shows that much of OSS development is driven by a small percentage of individuals. This phenomenon, known as ``participation inequality''~\cite{participant-inequality-in-oss}, highlights the disparity in contribution levels, with most work concentrated among a few core contributors. Mature OSS communities, such as the Apache Software Foundation,\footnote{https://www.apache.org/} are known to be inclusive \cite{asf-diversity-gender} but have been plagued by the same issue \cite{inequality-oss-asf}.

This study dives into a relatively modern and rapidly growing community to understand the inequality phenomenon in code contributors and the demographic distributions among them. The Rust programming language community exemplifies both the collaborative potential and sustainability challenges inherent in OSS. Rust has gained prominence within the OSS world, offering a unique blend of safety, performance, and memory control that appeals to industry and individual developers. Rust's focus on preventing memory safety issues and addressing challenges inherent in low-level programming languages traditionally perceived as less accessible due to complexity, often found in languages like C/C++, has contributed to its rising adoption~\cite{common-bug-fix-pattterns-rust}. In fact, Rust has been ranked the ``Most Admired'' programming language for eight consecutive years in the Stack Overflow Developer Survey.\footnote{\url{https://survey.stackoverflow.co/2024/}}

However, despite Rust's growing reputation, there remains limited understanding of the diversity within its contributor base and how well it reflects Rust's user community. It would also be worthwhile to understand how an ecosystem that is only a decade old is experiencing the challenges modern OSS projects face~\cite{why-oss-fail}. A key approach to understanding contributor dynamics in the ecosystem is exploring where the most technical interactions occur. In most OSS projects, PRs are one of the best ways to analyze collaboration~\cite{prs-better-than-commits}, where contributors submit proposed code changes, and reviewers assess and approve them before they are merged into the codebase. These interactions between \textbf{authors} (who submit PRs) and \textbf{reviewers} (who evaluate PRs) are central to the success of collaborative projects. 
Beyond technical contributions, understanding the relationships, power dynamics, and collaboration patterns within such interactions is essential for analyzing the inclusivity and sustainability of the Rust ecosystem.

Like many other open source software (OSS) projects, the Rust community is structured around a formal membership system~\cite{toral2010analysis}. Sustained contributions to core Rust projects are crucial for the language's evolution, and among these contributors, full members of the Rust GitHub organization occupy influential roles by virtue of their privileged access and responsibilities, such as voting on Requests for Comments (RFCs), which help shape the project's strategic direction.\footnote{\url{https://forge.rust-lang.org/compiler/membership.html}} Despite their importance, the extent to which these contributors mirror the global diversity of Rust's broader user base remains underexplored. This potential misalignment could significantly affect the community's ability to foster an inclusive environment that attracts and retains a diverse range of contributors worldwide.

To address this gap, this study aims to understand the diversity, collaboration patterns, and the role that core developers and sub-communities play in sustaining the Rust project. Specifically, we focus on answering the following questions: 

\begin{itemize}
\item \textbf{RQ1: How accurately does the diversity of active contributors reflect the broader Rust community's demographic composition?} 
\item \textbf{RQ2: What are the collaboration patterns and interaction dynamics among Rust contributors?} 
\item \textbf{RQ3: To what extent do subcommunities and top contributors within the Rust ecosystem exhibit diversity?} 
\end{itemize}

To answer these research questions, we focused on the core Rust compiler, Rust Analyzer (an integrated development environment (IDE) front-end), and Cargo (the Rust package manager). We examined the contributions and interactions within these repositories. 
Utilizing the GitHub API, we extracted PRs, authors, and reviewers. Then, we enriched this dataset with demographic information such as geographic location and likely gender. We then constructed social network graphs in R, applying the Louvain community detection algorithm to identify subcommunities within the network. Diversity within these subcommunities was quantitatively assessed using Blau's Index. At the same time, various network metrics were calculated to evaluate the structure and dynamics of Rust contributor interactions.

Our findings revealed that the diversity of active contributors does not fully reflect the broader Rust user community's demographic composition; this can be viewed in our anonymized replication that is publicly available  \cite{reproducibility}. The contributor base is heavily concentrated in specific geographic regions, with a predominance of contributors from the United States and Germany. Gender diversity remains extremely low, with men contributors dominating participation. Additionally, non-member contributors play a crucial role in the collaboration networks, often forming distinct clusters independent of official Rust organization members. Subcommunities demonstrate substantial geographic diversity; however, gender diversity remains notably low across all groups. These findings suggest that 
the community may need more inclusive practices to attract a broader range of participants to reflect its user base as it leads to more adoption, which in turn leads to more job creation that benefits the whole ecosystem \cite{carter2021diversity-linux}, aligns with Rust's ethos of empowering everyone and capturing contributors from upcoming communities in Southern and Eastern parts of the world \cite{github-survey-2023}.

The paper is organized as follows: Section \ref{related-work} reviews related work, Section \ref{methodology} explains the study's methods, and Section \ref{results} presents findings on contributor demographics and network structures. Section \ref{discussion} addresses the implications, while Section \ref{limitations-future-work} lists the limitations and future directions. Finally, Section \ref{conclusion} summarizes our contributions.

\section{Related work}
\label{related-work}

The success and sustainability of OSS projects have been observed to correlate with their contributor communities' collaborative efforts and network structures. Previous research has focused on understanding these dynamics, particularly through the lens of social network analysis and diversity studies.

\subsection{Social Network Analysis in OSS Communities}

Social network analysis has been a pivotal tool in dissecting the collaborative structures within OSS communities. Liu and Gui~\cite{collab-networks-oss} conducted a structural analysis of collaboration networks across three OSS communities, revealing that these networks predominantly exhibit a modular small-world structure. The study distinguished between single-dimensional developers, who focus on specific sub-projects, and multi-dimensional developers, who engage across multiple modules, highlighting the diverse collaboration patterns. 

A comprehensive review categorized existing research on OSS social networks into Structure, Lifecycle, and Communication~\cite{categorization-network-science}. The review emphasized that project structure significantly influences success, with factors such as a structured hierarchy, diverse developer base, and project prominence being critical. However, the authors noted a gap in temporal analyses, suggesting the need for studies that track network evolution over time to understand sustainability factors better.

Banks et al.~\cite{python-network-science} focused on Python packages hosted on GitHub, employing network-based metrics like degree and eigenvector centrality to measure OSS innovation and impact. Their findings demonstrated a positive correlation between centrality measures and package downloads, indicating that more central packages within the network tend to achieve higher usage and visibility. 

Yin et al.~\cite{socio-technical-networks-institute-analysis} integrated institutional analysis with socio-technical networks to assess the health of OSS projects within the Apache Software Foundation (ASF) incubator. By combining governance structures with collaboration networks, the study provided a holistic view of factors contributing to project sustainability, emphasizing the interplay between institutional policies and developer interactions.

An empirical analysis of GitHub project social team structures reveals a significant correlation between the speed of pull-request processing and the density of the collaboration network 
\cite{empirical-social-teams-github}. The study found that smaller to medium-sized projects benefit from a tightly-knit core of maintainers who engage in repeated interactions, facilitating more efficient review and integration of contributions. Building on these findings, we aim to investigate similar dynamics within three significant repositories in the Rust ecosystem. Our analysis will examine the network structures between Rust Github organization members (typically those with privileged permissions and approval access) and non-members, further exploring how geographic and gender demographics influence interaction patterns.

While significant strides have been made in understanding network structures and diversity within OSS communities, there remains a lack of research that integrates these dimensions to explore how demographic factors intersect with collaboration patterns. Most existing studies either focus on network analysis without incorporating demographic attributes or examine diversity without getting into the underlying social dynamics.

\subsection{Diversity in OSS}
Research on diversity within OSS communities has garnered increasing attention due to its potential impact on participation, engagement, and project outcomes~\cite{rodriguez2021perceived,bosu2019diversity,nadri2021relationshipdevelopersperceptiblerace}. Shiek et al. \cite{collaborative-groups-analysis} conducted an empirical evaluation on the relationship between the racial and ethnic diversity of OSS groups and their frequency of contributions. They found that heterogeneous groups tend to have a higher median number of contributions than homogeneous groups.

The Apache Software Foundation (ASF) conducted a diversity survey \cite{asf-survey-gender-diversity} that highlighted ongoing challenges related to gender and geographic diversity within their projects. Despite being well-established, ASF projects struggle to attract and retain a diverse pool of contributors, underscoring the need for targeted diversity and inclusion initiatives.

A recent study~\cite{prana2021including} explores gender diversity within OSS projects hosted on GitHub, focusing on geographic disparities. Their findings show that gender diversity remains low across all regions, with only slight variations between them.

\subsection{Rust Community}
Yuxia et al. \cite{free-paid-devs-rust-ecosystem} investigated the dynamics between paid developers and volunteer contributors. The study highlighted the nuanced relationships and dependencies between different types of contributors, suggesting that organizations supporting OSS projects must adopt sensitive integration strategies to balance these dynamics effectively.

Researchers have studied a framework for assessing systemic risks in software ecosystems~\cite{schueller2022modelinginterconnectedsocialtechnical} by analyzing dependency networks and developer activity. Their analysis of the Rust ecosystem revealed that key developers often maintain multiple libraries, posing potential risks if these individuals become inactive. William Schueller et al. emphasize the critical role of understanding the interconnectedness of social and technical factors in ensuring the resilience of OSS projects.

Recent works have used network analysis \cite{imtiaz2023trustingcodewildsocial-pre-print} in the Rust ecosystem, constructing developer networks based on collaboration activities. The study proposed integrating contributor reputation metrics to enhance dependency management practices, highlighting the importance of social factors in technical decision-making processes within OSS communities.

Our study seeks to bridge the current state of research done in the Rust ecosystem gap by explicitly analyzing the social dynamics and demographic diversity within the Rust community. By integrating social network analysis with demographic data such as geographic location and gender, we aim to uncover how diversity manifests within subcommunities and influences collaboration patterns.
Unlike previous studies that have focused solely on centrality measures or diversity aspects, our research provides a comprehensive view combining both elements, offering deeper insights into the factors that drive participation and collaboration in the Rust ecosystem.






\begin{figure}[htbp]
\centering
\includegraphics[width=0.7\linewidth]{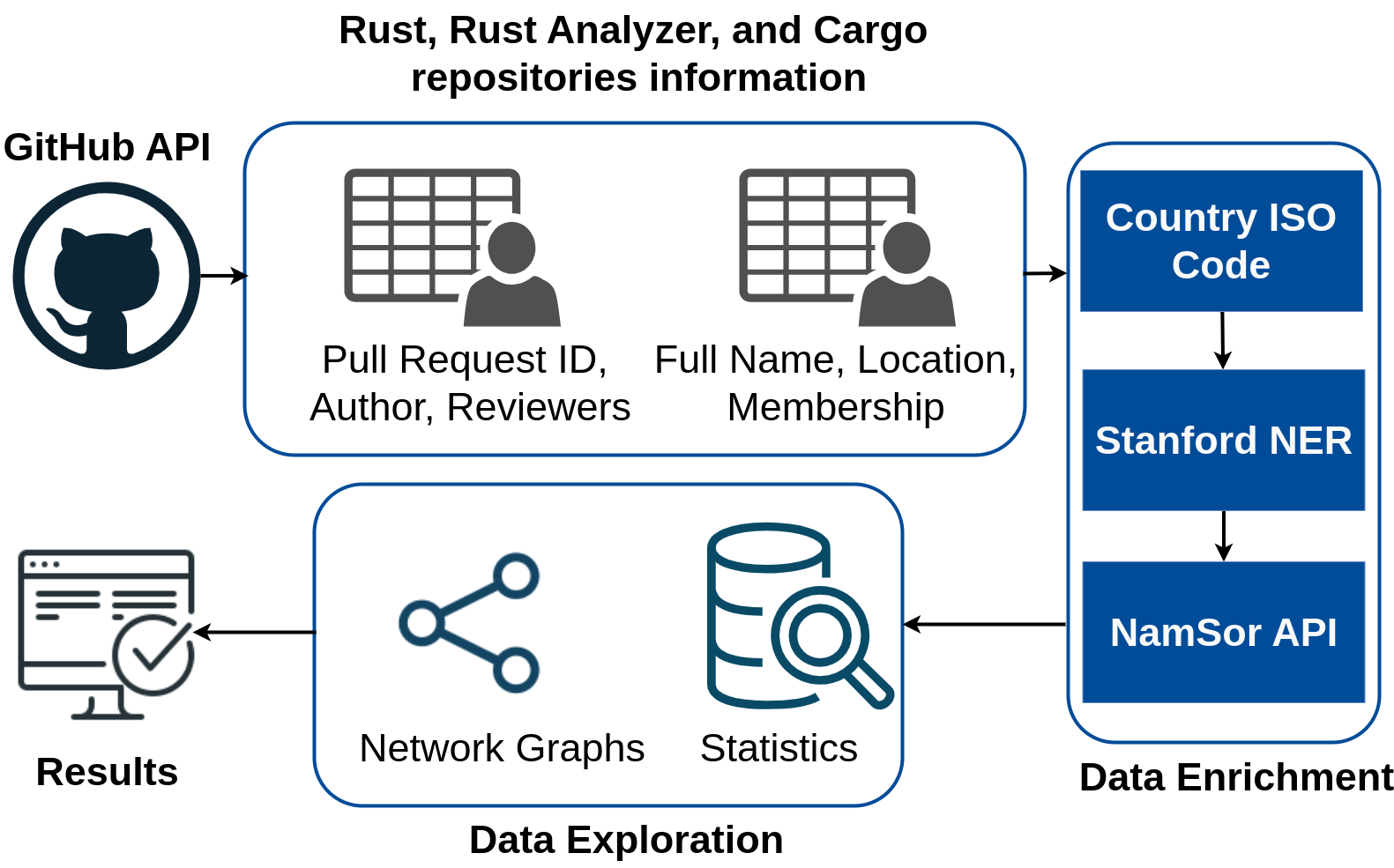}
\caption{Methodology overview used to conduct research}
\label{fig:methodology-flowchart}
\end{figure}

\section{Methodology}
\label{methodology}

The methodology employed in this research is illustrated in Figure \ref{fig:methodology-flowchart}, providing a clear and sequential overview of our approach. 

\subsection{Projects Selection Criteria}
As of January 2025, the Rust organization on GitHub comprises approximately 222 repositories. To focus our analysis on the most impactful projects, we selected the top three repositories based on the number of stars, a common indicator of popularity and community engagement \cite{large-sclae-study-why-stars-github-popularity}. We excluded repositories, such as \textit{book}, \textit{mdBook}, and \textit{rustlings}, as they are intended for documentation and learning purposes. After this refinement, our study concentrated on three core repositories:

\begin{enumerate}
    \item \textbf{Rust (101k+ stars)}: The core Rust compiler repository \cite{rust-lang-rust}. It has around 5374 lifetime contributors and is the largest repository in the Rust ecosystem.
    \item \textbf{Rust Analyzer (14.6k stars)}: A Rust compiler front-end designed for integrated development environments (IDEs) \cite{rust-analyzer}. A total of 918 lifetime contributors are empowered by backing from companies such as Ferrous Systems.\footnote{https://ferrous-systems.com/}
    \item \textbf{Cargo (13.1k stars)}: The official package manager for Rust, responsible for managing project dependencies and builds \cite{rust-lang-cargo}. Comprising a cumulative total of 1053 contributors over its lifetime.
\end{enumerate}

Furthermore, these three repositories represent distinct facets of the Rust ecosystem, each demanding unique skill sets and domain expertise. The \textbf{Rust} compiler involves low-level systems programming and compiler design, requiring contributors to understand language internals and performance optimization deeply. \textbf{Rust Analyzer} focuses on IDE integration and developer tooling, necessitating knowledge in language server protocols, code analysis, and user experience considerations. \textbf{Cargo}, as the package manager, revolves around dependency management and building systems, where contributors need proficiency in software distribution and project configuration. Each project follows unique development and collaboration practices tailored to the specific goals and challenges mentioned in their dedicated guidelines.\footnote{\url{https://rustc-dev-guide.rust-lang.org/contributing.html}}\footnote{\url{https://github.com/rust-lang/rust-analyzer/blob/master/docs/dev/guide.md}}\footnote{\url{https://doc.crates.io/contrib/}}  By including these repositories, our study captures a broad range of contribution dynamics and interactions within the Rust community, providing insights into how varying technical domains influence participation and collaboration dynamics. We classify contributors into two categories: \emph{members}, accounts that belong to the public \texttt{rust-lang} GitHub organization and therefore have merge and branch-protection privileges, and \emph{non-members}, who contribute without those elevated rights.

\subsection{Extracting PRs, Author, and Reviewers}

We posit that individuals who invested the time to complete the comprehensive Rust Annual Community Survey 2023 \cite{rust-annual-survey-2023}, which takes approximately 25 minutes, represent the community's distribution due to their significant level of engagement and commitment. Hence, to understand and compare the contributors' dynamics preceding the Rust Annual Survey 2023, we collected PR data from the year prior to the survey's release, specifically from December 18th, 2022, to December 18th, 2023. 

We used the GitHub REST API to extract the usernames of PR authors and reviewers. However, in the Rust ecosystem, the review and approval process often involves a bot named \textit{Bors}, which automates the merging of approved PRs based on comments within the discussion thread. Because approvals managed by \textit{Bors} may not be fully captured through standard API endpoints that fetch only GitHub assigned reviewers, to get past that, we used a custom regex script that identifies the approval text in the PR comments and extracts the usernames of approving reviewers that are not directly accessible. 

Additionally, we excluded PRs involving automated bots such as \textit{dependabot}, \textit{renovatebot}, and \textit{rustbot} to ensure our analysis focused solely on human contributors. 

\subsection{Enriching Data with Full Name, Location, and Membership}

After compiling the list of usernames, we gathered additional profile information for each contributor. This information included their full name, location, and membership status within the Rust GitHub organization. We enriched the dataset by creating one-to-one PR pairs, mapping each PR author to each of their respective reviewers individually. This approach allowed us to construct a comprehensive contributors' dataset, incrementing the contribution count for every instance a user authored or reviewed a PR. 

To ascertain the geographical distribution of contributors, we analyzed the location details provided in user profiles. We utilized the \texttt{geograpy3}\footnote{\url{https://pypi.org/project/geograpy3/}} Python package, which leverages the Natural Language Toolkit (NLTK) framework, to parse the city or location strings and extract corresponding country information. The extracted data were then converted to ISO2 country codes\footnote{\url{https://www.iso.org/obp/ui/\#search}} using the \texttt{country\_converter} package. To ensure the highest accuracy, we manually verified the location data by searching the location details (city, country emoji, etc) online with the assigned country code, correcting any inconsistencies or errors identified during the automated extraction process. Overall, about a third of the data was inconsistently classified. For example, ``Earth'' and ``UK'' were classified under the ``US'' country code.

\begin{table*}[ht]
\begin{threeparttable}
\setlength{\tabcolsep}{2pt} 
\centering
\caption{Statistics of code contributors' demographics in three crucial Rust projects\tnote{†}}
\begin{scriptsize}
\normalsize
\begin{tabular}{@{}lccccccc@{}}
\textbf{Country} & \textbf{Men \%} & \textbf{Women \%} & \textbf{Unknown \%} & \textbf{Member (\%)} & \textbf{Non-Member (\%)} & \textbf{Contribution Share (\%)\tnote{1}} & \textbf{Contributor Share (\%) \tnote{2}} \\
\hline \hline
\multicolumn{8}{c}{\textbf{Rust Project (Domain: Compiler development, Rust language internals, LLVM, Systems programming)}} \\
\hline
USA & 86 & 6  & 8  & 16 & \textbf{84} & 30 & 16 \\
Germany                 & 88 & 3  & 9  & 19 & \textbf{81} & 13 & 7  \\
France                  & 75 & 8  & 17 & 13 & 88 & 5  & 2  \\
United Kingdom          & 83 & 2  & 14 & 12 & 88 & 3  & 4  \\
Switzerland             & 86 & \textbf{14} & 0  & 14 & 86 & 2  & 1  \\
Other countries         & 80 & 8  & 12 & 10 & 90 & 14 & 23 \\
Unidentified            & 0  & 0  & 100 & 8  & 92 & 33 & 47 \\
\hline
\multicolumn{8}{c}{\textbf{Rust Analyzer Project (Domain: Language server protocol, Syntax parsing, Editor tooling)}} \\
\hline
Germany                 & 94 & 0   & 6  & 17 & 83 & 46 & 9  \\
Romania                 & 100 & 0  & 0  & 100 & 0   & 14 & \textbf{0}  \\
Iran                    & 0   & 0  & 100 & 100 & 0   & 11 & \textbf{0}  \\
Japan                   & 71  & \textbf{14} & 14 & 0   & 100 & 6  & \textbf{3}  \\
USA & 84 & 10 & 6  & 6  & 94  & 4  & 15 \\
Other countries         & 83 & 6  & 11 & 10 & 90 & 8  & 34 \\
Unidentified            & 0  & 0  & 100 & 7  & 93  & 10 & 39 \\
\hline
\multicolumn{8}{c}{\textbf{Cargo Project (Domain: Build systems, Project configuration, package management)}} \\
\hline
USA & 81 & 6  & 13 & 16 & 84 & 34 & 15 \\
Taiwan                  & 67 & 0  & 33 & 33 & 67 & 30 & 1  \\
Germany                 & 92 & 8  & 0  & 33 & 67 & 2  & 6  \\
China & 78 & \textbf{22} & 0  & 0   & \textbf{100} & 1  & \textbf{4}  \\
United Kingdom          & 90 & 10 & 0  & 10  & 90  & 1  & 5  \\
Other countries         & 89 & 3  & 8  & 10 & 90 & 4  & 30 \\
Unidentified            & 0  & 0  & 100 & 19 & 81 & 28 & 39 \\
\hline

\label{tab:community-stats}
\end{tabular}
 \begin{tablenotes}
    \item[1] The percentage of contributions made by users from a specific country divided by the total contributions made by users in the repository.
    \item[2] The percentage of users from the total number of individuals involved in the PR process that belong to a particular country.
    \item[†] Values shown in \textbf{bold} show higher likely women participation, full non-member participation, and low contributor share.
  \end{tablenotes}
\end{scriptsize}
\end{threeparttable}
\end{table*}

For gender identification, we relied on contributors' full names. We employed Stanford's Named Entity Recognition (NER) tool, available through the Python package \texttt{stanza}\footnote{\url{https://stanfordnlp.github.io/stanza/}} to determine whether the extracted names referred to persons. This filtered out usernames that might represent organizations, projects, or non-person entities. Only names recognized as ``PERSON'' entities, with additional manual verification, were considered in subsequent analyses. Subsequently, we used the NamSor API, a widely recognized name classification tool in academic research \cite{namsor-gender-comparitive,namsor-usage-paper}, to predict the likely gender of each contributor. By inputting the full name and corresponding ISO2 country code into NamSor, we obtained gender probabilities for each individual. To maintain a high level of confidence in our gender assignments, we included only those contributors for whom NamSor provided a gender probability of 80\% or higher. Any cases with lower confidence levels or potential false positives underwent additional manual verification. Both authors independently reviewed these ambiguous cases, and only upon mutual agreement were the final gender classifications assigned. 

\subsection{Social Network Analysis}
\label{social-network-analysis-methodology}
Post enriched data collection, we constructed social network graphs using the R programming language and the \texttt{igraph}\footnote{\url{https://igraph.org/}} package. Each contributor is represented as a \emph{node}, and the \emph{edges} represent user interactions in these graphs. We developed two types of graphs: directed and undirected. In the directed graphs, where edges from \emph{authors} to \emph{reviewers} represent reviewed pull requests (PRs), and undirected graphs, where edges denote PR-based interactions. The former captured reciprocity and influence, while the latter revealed community structures, offering a holistic view of contributor dynamics.

To identify clusters within the social network, we applied the \emph{Louvain community detection algorithm}, which efficiently uncovers communities by optimizing modularity \cite{louvain-paper}. This algorithm allowed us to detect subcommunities where contributors are more densely connected among themselves than with the rest of the network. Incorporating demographic details into our analysis enabled us to understand various collaboration dynamics. We examined how \textbf{members} and \textbf{non-members} of the Rust GitHub organization collaborate, exploring whether official affiliation influences interaction patterns. Additionally, we assessed the diversity within sub-communities by considering contributors' geographic locations (using ISO2 country codes) and likely gender. This approach provided insights into how contributors from different backgrounds interact within the network. 

The graphs were constructed by including nodes only when the respective node demographic information was available. To ensure the accuracy and relevance of our network analysis, we removed \emph{self-loops}, which are edges that connect nodes to themselves, as they do not represent meaningful collaborations between different contributors. Furthermore, we excluded \emph{isolated nodes} that remained disconnected after the removal of self-loops, focusing our analysis on the interconnected portion of the network where actual interactions occur. 

We calculated various network metrics~\cite{newman2018networks} to quantify the properties of the collaboration networks. \emph{Reciprocity} was measured to understand the extent of mutual connections between contributors, indicating how often interactions are bidirectional. \emph{Network density} was analyzed with respect to contributors' locations and membership status to assess whether collaborations are highly concentrated or distributed in a manner resembling real-world social networks. 

To capture subcommunity geographic diversity, we constructed graphs using ISO2 country codes and used \emph{Blau's Index}\cite{blau1977inequality} to calculate the diversity quantitatively in identified subcommunities across the projects \cite{collaborative-groups-analysis}. 

Blau's Index is defined as:

\begin{equation}
\text{Blau's Index} = 1 - \sum_{i=1}^{k} p_i^2
\end{equation}
\noindent where:
\begin{itemize}
    \item \( k \) is the number of distinct categories (e.g., geographic location groups).
    \item \( p_i \) represents the proportion of individuals in category \( i \) within the group.
\end{itemize}

A higher Blau's Index value indicates greater diversity within the subcommunity, while a lower value suggests homogeneity. By applying this metric to the geographic distributions of contributors, we could quantitatively compare the diversity levels across different subcommunities within the Rust ecosystem. Additionally, we employed \emph{centrality measures}~\cite{newman2018networks}, such as degree centrality (the number of direct connections a node has), betweenness centrality (the extent to which a node lies on paths between other nodes), and eigenvector centrality (the influence of a node based on the importance of its connections), to identify influential contributors within the network. By examining the demographics of these central contributors, we aimed to understand which user groups hold significant positions in the collaboration network and how their influence might impact the overall dynamics. This network analysis provided a comprehensive view of the collaboration patterns among contributors, highlighting the roles of different demographic groups and the structure of the contributor community within the Rust ecosystem.

\section{Results}
\label{results}
\subsection{RQ1: How accurately does the diversity of active contributors reflect the broader Rust community's demographic composition?}
\label{rq1}


Table~\ref{tab:community-stats} presents a snapshot of the demographics of the contributors in the three codebases considered in the Rust ecosystem grouped with respect to their country, organized in descending order based on their \emph{\textbf{Contribution Share (\%)}} to each repository, which is calculated based on the contributions made by users from a specific country by the total contributions made by users to the repository. Similarly, \emph{\textbf{Contributor Share (\%)}} is the percentage of contributors from the total number of individuals involved in the PR process over the last year that belong to a particular country. We later spread these grouped country contributions on identified demographic data of the user such as their membership status in Rust GitHub organization and likely calculated Gender. Table~\ref{tab:community-stats} includes the top five countries in terms of contributions for each project, followed by grouping all other countries under "Other countries" and contributors whose country of origin could not be identified under "Unidentified."

Overall, we identified contributors from 50 countries in the Rust project, 36 in Rust Analyzer, and 30 in Cargo, with 55 unique countries across three projects. The distribution of top contributing countries varied across the projects:

\textbf{Rust Project}: Predominantly, contributions came from the United States (US) (30.16\% contribution share, 16.25\% contributor share) and Germany (12.61\%, 6.71\%), followed by France, United Kingdom, and Switzerland, indicating a concentration in North America and Western Europe. The "Unidentified" group accounted for 32.89\% of contributions and had the highest contributor share at 47.08\%.

\textbf{Rust Analyzer Project}: Germany led with a substantial 46.03\% contribution share but a lower contributor share of 8.53\%, indicating significant contributions from a small number of individuals. Romania and Iran also had high contribution shares (13.96\% and 11.24\%, respectively) with minimal contributor shares (both at 0.47\%), further emphasizing concentrated contributions. The "Unidentified" category contributed 10.34\% with a contributor share of 38.86\%.

\textbf{Cargo Project}: The US again topped the list with a 34.48\% contribution share and a 15.05\% contributor share. Taiwan had a notable 29.66\% contribution share but a low contributor share of 1.46\%, indicating that few contributors from Taiwan made substantial contributions. The "Unidentified" contributors accounted for 27.76\% of contributions and had a contributor share of 38.83\%.

Gender representation among contributors was consistently skewed across all projects, with individuals identified as women comprising only 3.4\% in Rust, 3.7\% in Rust Analyzer, and 3.8\% in Cargo. These figures align with the survey's report of 3.5\% women participants \cite{rust-annual-survey-2023}. However, gender disparity varied by country and project:

\begin{itemize}
    \item In the \textbf{Rust project}, the US had a higher proportion of women contributors (5.99\%) compared to the overall average, while Germany had a lower representation (2.90\%). 
    \item In the \textbf{Rust Analyzer project}, Germany had no identified women contributors despite being the top contributing country. Japan had a higher women participation rate at 14.29\%.
    \item In the \textbf{Cargo project}, China had a significantly higher proportion of women contributors at 22.22\%.
\end{itemize}

Regarding membership status within the Rust GitHub organization, non-members constituted the majority of contributors across all countries and projects. In most cases, over 80\% of contributors were non-members.

The \emph{Contribution Share (\%)} of the top country in each project was higher than the combined contribution shares of the next four countries. For example, in the Rust Analyzer project, Germany's contribution share (46.03\%) exceeded that of Romania, Iran, Japan, and the US combined. Similarly, the \emph{Contributor Share (\%)} of the top country was generally higher than that of the following four countries combined, except for the US in the Rust Analyzer project.

Comparatively, the community distribution according to the 2023 Rust Annual Survey \cite{rust-annual-survey-2023}, the top ten countries represented among participants were the United States (22\%), Germany (12\%), China (6\%), the United Kingdom (6\%), France (6\%), Canada (3\%), Russia (3\%), the Netherlands (3\%), Japan (3\%), and Poland (3\%), with participants hailing from approximately 100 countries.

\begin{figure}[htbp]
  \centering
  \includegraphics[width=0.9\columnwidth]{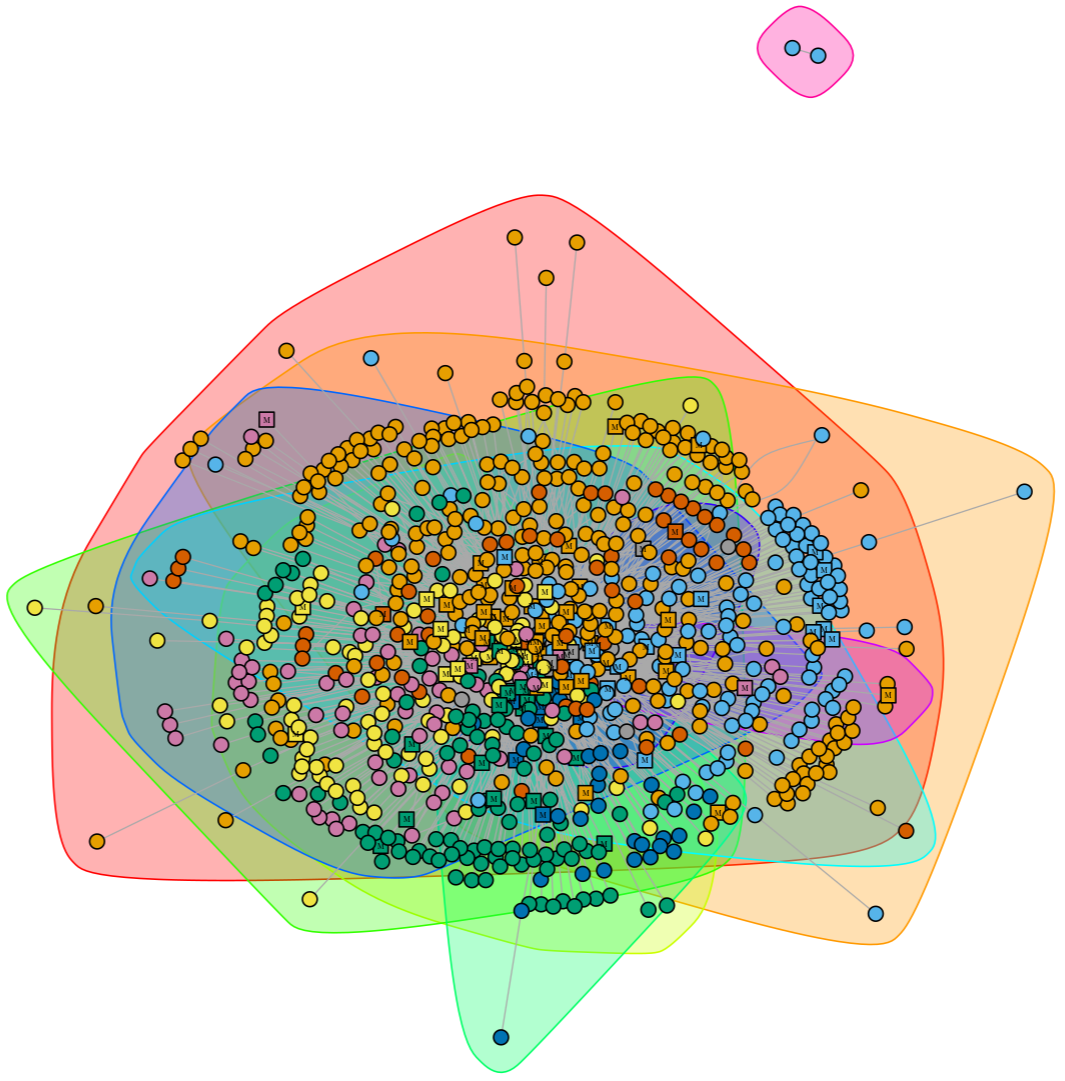}
  \caption{Rust project contributors' social network (A larger image is available in our replication package)}
  \label{fig:rust-project-graph-member}
\end{figure}

\subsection{RQ2: What are the collaboration patterns and interaction dynamics among Rust contributors?}
\label{rq2}
\begin{figure}[htbp]
  \centering
  \includegraphics[width=0.6\columnwidth]{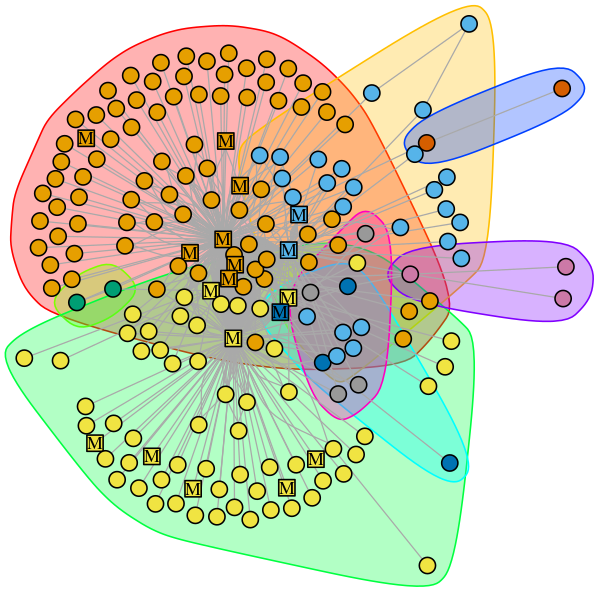}
  \caption{Rust Analyzer Project contributors' social network}
  \label{fig:rust-analyzer-graph-member}
\end{figure}

\begin{figure}[htbp]
  \centering
  \includegraphics[width=0.6\linewidth]{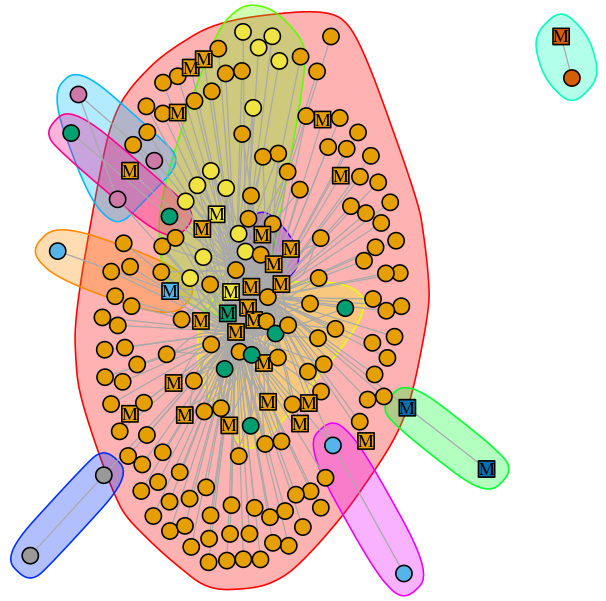}
  \caption{Cargo Project contributors' social network}
  \label{fig:cargo-network-member}
\end{figure}

We constructed undirected social network graphs with membership data for three projects: Rust, Rust Analyzer, and Cargo. These graphs are depicted in Figure~\ref{fig:rust-project-graph-member}, Figure~\ref{fig:rust-analyzer-graph-member}, and Figure~\ref{fig:cargo-network-member}, respectively.

In these Figures, contributors are represented as nodes, with \ding{108} circles denoting \textbf{non-members} and \ding{110} squares labeled with an "M" indicating \textbf{members} of the Rust GitHub organization. Edges between nodes represent collaboration through PRs, where an interaction has occurred between the author and the reviewer. Analysis of the network graphs reveals distinct clustering patterns colored in different colors, depicting contributor groups formed within each project. 

In the Rust project, one of the clusters does not contain any member nodes. Similarly, in both the Rust Analyzer and Cargo projects, there are three clusters each that lack member nodes. Member nodes are predominantly clustered near the center of the networks in the Rust project. In contrast, the Cargo and Rust Analyzer project's member nodes are more dispersed. 

The structural properties of these networks were further examined through the calculation of various graph metrics, as presented in Table~\ref{tab:network-collab-metrics}. The \emph{reciprocity} of the directed versions of these networks provides insight into the mutual nature of collaborations. Reciprocity measures the likelihood of node pairs having reciprocal connections, in this context, whether contributors review and submit PRs to each other. The Cargo project exhibited a reciprocity score of 0.5, indicating a relatively balanced two-way collaboration among contributors. In contrast, the Rust and Rust Analyzer projects showed lower reciprocity scores. All the undirected networks, constructed only using ISO Codes and member status nodes separately, displayed low \emph{density} values, which is characteristic of real-world social networks and OSS networks in particular \cite{empirical-social-teams-github}.

\begin{table}[ht]
\centering
\caption{Network Metrics for Rust Projects}
\label{tab:network-collab-metrics}
\begin{tabular}{l c c c}
\textbf{Metric}        & \textbf{Rust} & \textbf{Rust Analyzer} & \textbf{Cargo} \\
\hline
\hline
Reciprocity            & 0.26          & 0.15                   & \textbf{0.50 }          \\
\hline
Density (Member)       & 0.02          & 0.04                   & 0.06           \\
\hline
Density (Country ISO Code)     & 0.05          & 0.08                   & 0.07           \\
\hline
\end{tabular}
\end{table}


\subsection{RQ3: To what extent do subcommunities and top contributors within the Rust ecosystem exhibit diversity?}

We analyzed the subcommunities formed within the social undirected graphs of the Rust, Rust Analyzer, and Cargo projects. The diversity within these subcommunities was quantified using Blau's Index, as presented in Table~\ref{tab:blau-index}. Eight subcommunities were identified in the \textbf{Rust project}, with Blau's Index scores ranging from 0.000 to 0.883. Most of these communities exhibited high geographical diversity, with Blau's Index scores above 0.85, indicating a diverse mix of contributors from different countries. Community 1 and Community 2 had Blau's Index scores of 0.857 and 0.883, respectively, reflecting substantial diversity. An exception was Community 8, which had a Blau's Index of 0.000 due to comprising only two contributors from the same country, making it an outlier. 

Two subcommunities were formed for the \textbf{Rust Analyzer project}, displaying high geographical diversity with Blau's Index scores of 0.889 and 0.895. This result suggests that contributors within these communities are from a wide range of countries, contributing to a diverse collaborative environment. 

In the \textbf{Cargo project}, six subcommunities were identified. Community 1 had a high Blau's Index of 0.914, indicating significant geographical diversity. The remaining five communities each had a Blau's Index of 0.500, suggesting moderate diversity within these groups. 


\begin{table}[ht]
\centering
\caption{Blau's Index for Communities in Rust Projects}
\begin{tabular}{llc}
\textbf{Project} & \textbf{Community} & \textbf{Blau's Index} \\
\hline
\hline
\multirow{8}{*}{Rust}
 & Community 1 & 0.857 \\
 & Community 2 & 0.883 \\
 & Community 3 & 0.852 \\
 & Community 4 & 0.852 \\
 & Community 5 & 0.861 \\
 & Community 6 & 0.762 \\
 & Community 7 & 0.720 \\
 & Community 8 & 0.000 \\
\hline
\multirow{2}{*}{Rust Analyzer}
 & Community 1 & 0.889 \\
 & Community 2 & 0.895 \\
\hline
\multirow{6}{*}{Cargo}
 & Community 1 & 0.914 \\
 & Community 2 & 0.500 \\
 & Community 3 & 0.500 \\
 & Community 4 & 0.500 \\
 & Community 5 & 0.500 \\
 & Community 6 & 0.500 \\
\hline
\label{tab:blau-index}
\end{tabular}
\end{table}

To understand the influence and roles of influential contributors across subcommunities, we calculated centrality measures, degree centrality, betweenness centrality, and eigenvector centrality, for the contributors across the projects, as shown in Table \ref{tab:centrality_scores_grouped}. Across all projects, the top contributors identified through centrality measures were predominantly men, with the \emph{Likely Gender} being \emph{men} or \emph{Unknown}. There were no women contributors among the most central nodes. 
The users with the most edges connected to them, indicating top contributions via \emph{degree centrality} metrics, are shown in Table \ref{tab:centrality_scores_grouped}. Geographically, in the \textbf{Rust project}, they were primarily from the United States (US), Germany (DE), and France (FR). In the \textbf{Rust Analyzer project}, top contributors hailed from Germany (DE), Romania (RO), Japan (JP), and Iran (IR), indicating a broader geographical spread compared to the Rust project. The \textbf{Cargo project} featured top contributors from Taiwan (TW), the United States (US), and Germany (DE). The same geographic distribution was seen in the third part of \ref{tab:centrality_scores_grouped}, representing users connected to the most influential nodes in the respective project networks, captured through \emph{eigenvector centrality}.

The middle part of Table \ref{tab:centrality_scores_grouped} depicts the top users with the highest contributions going through them (node) in the network, representing the best nodes for the fastest flow of contributions. Both members and non-members of the Rust GitHub organization were among the top contributors. For instance, in the Rust project, one of the top contributors with a high degree of centrality was a non-member. This suggests that significant contributions and influence within the projects are not confined to official members. The \emph{betweenness centrality} scores varied across the projects. In the Rust project, the top betweenness centrality scores ranged from 0.10 to 0.26, indicating a less centralized network where contributions do not predominantly pass through a few individuals. Conversely, the Rust Analyzer project exhibited higher betweenness centrality scores, with the top contributor scoring 0.72, suggesting that specific contributors play critical roles as this indicates many contributions (PR authorship or reviews) go through them from/to different parts of the network acting as a central piece. Similarly, the top betweenness centrality scores in the Cargo project were 0.77 and 0.48, indicating a more centralized network structure.

\begin{table*}[ht]
\begin{threeparttable}
\setlength{\tabcolsep}{6pt}  
\centering
\caption{Centrality scores of top contributors grouped by repository and centrality type\tnote{†}}
\label{tab:centrality_scores_grouped}
\begin{scriptsize}
\normalsize
\begin{tabular}{@{}l l r r r || r r r r || r r r r@{}}
 & \multicolumn{4}{c}{\textbf{Degree Centrality}} &
   \multicolumn{4}{c}{\textbf{Betweenness Centrality}} &
   \multicolumn{4}{c}{\textbf{Eigenvector Centrality}} \\ \hline\hline
\multicolumn{13}{c}{\textbf{Rust Project}} \\ \hline
\textbf{Rank} & \textbf{ISO} & \textbf{Gender} & \textbf{Score} & \textbf{Member} &
               \textbf{ISO} & \textbf{Gender} & \textbf{Score} & \textbf{Member} &
               \textbf{ISO} & \textbf{Gender} & \textbf{Score} & \textbf{Member} \\ \hline
1 & US & Men     & 1253 & Yes & US & Men     & 0.26 & Yes & US & Men     & 1.00 & Yes \\
2 & DE & Men     &  536 & Yes & US & Men     & 0.23 & Yes & DE & Unknown & 0.75 & Yes \\
3 & FR & Men     &  493 & \textbf{No}  & US & Men & 0.11 & \textbf{No} & FR & Men & 0.59 & No \\
4 & US & Men     &  412 & Yes & DE & Men     & 0.10 & Yes & US & Men     & 0.57 & Yes \\
5 & DE & Unknown &  390 & No  & US & Men     & 0.10 & No  & DE & Men     & 0.52 & No \\ \hline
\multicolumn{13}{c}{\textbf{Rust Analyzer Project}} \\ \hline
1 & DE & Men & 330 & Yes & DE & Men & 0.72 & Yes & DE & Men & 1.00 & Yes \\
2 & \textbf{RO} & Men & 174 & \textbf{No} & RO & Men & 0.49 & No & RO & Men & 0.68 & No \\
3 & \textbf{JP} & Unknown & 105 & \textbf{No} & \textbf{IR} & Unknown & 0.10 & No & JP & Unknown & 0.67 & No \\
4 & \textbf{IR} & Unknown & 101 & \textbf{No} & PT & Men & 0.04 & No & IR & Unknown & 0.57 & No \\
5 & DE & Men & 43 & Yes & US & Men & 0.01 & Yes & DE & Men & 0.30 & Yes \\ \hline
\multicolumn{13}{c}{\textbf{Cargo Project}} \\ \hline
1 & \textbf{TW} & Men & 371 & Yes & \textbf{TW} & Men & 0.77 & Yes & US & Men & 1.00 & Yes \\
2 & US & Men & 348 & Yes & US & Men & 0.48 & Yes & \textbf{TW} & Men & 0.99 & Yes \\
3 & US & Men &  52 & No  & US & Men & 0.04 & No  & US & Men & 0.18 & No \\
4 & US & Men &  41 & No  & US & Men & 0.04 & No  & US & Men & 0.14 & \textbf{No} \\
5 & DE & Men &  11 & No  & DK & Men & 0.02 & No  & DE & Men & 0.04 & \textbf{No} \\ \hline
\end{tabular}
\begin{tablenotes}
    \item[†] Few entries in \textbf{bold} depict contributors whose country or membership status are outliers.

\end{tablenotes}
\end{scriptsize}
\end{threeparttable}
\end{table*}

\section{Discussion}
\label{discussion}
\subsection{Contributors vs. Community Demographics}

\textbf{Geographic Disparities}: While the survey indicates any form of user engagement or participation from around 100 countries, our findings from RQ1 indicate that active code contributions are concentrated in fewer countries across the studied projects. The Rust project shows contributions from 50 countries, Rust Analyzer from 36, and Cargo from 30, with 55 unique countries across projects. This suggests that active contributions are more concentrated while community engagement is widespread. The United States and Germany are prominent in both the survey and contributor data, signifying strong representation. However, some countries like China, which ranks third in the survey (6\% of participants), have minimal representation among top contributors, appearing only in the Cargo project with a modest contribution share (0.93\%). This finding aligns with previous research on OSS participation imbalances~\cite{gasparini2020participation}, such as the 90-9-1 rule, which highlights that 90\% of users are passive observers, 9\% contribute occasionally, and only 1\%, the ``superusers'', generate the majority of contributions~\cite{nielsen2006}. Conversely, countries such as Romania and Iran, which are not highlighted in the survey's top ten, have substantial contribution shares in the Rust Analyzer project (13.96\% and 11.24\%, respectively), indicating that certain regions may have a small number of highly active contributors despite lower overall community participation. 

\textbf{Gender Disparities}: It remains a critical concern as previous studies have also shown~\cite{trinkenreich2020hidden,bosu2019diversity}. The low percentage of women contributors across all projects mirrors the survey's findings, highlighting persistent gender disparity. Notably, some countries exhibit higher women participation rates; for instance, China in the Cargo project has 22.22\% women contributors, significantly above the overall average. This variation suggests that regional factors may influence gender representation in contributions.

\textbf{Anonymous Contributors:} The high proportion of contributors with ``Unknown'' gender and ``Unidentified'' country underscores limitations in data collection and the challenges of accurately assessing diversity based on publicly available information. This could imply that individuals from traditionally underrepresented groups may feel compelled to conceal their identity to contribute effectively without encountering resistance \cite{nadri2021relationshipdevelopersperceptiblerace}.

\textbf{Membership Disparities}: The predominance of non-member contributors suggests that the Rust ecosystem comprises a huge number of external participants. While this reflects the open nature of OSS, it may also point to opportunities for better integrating these contributors into the formal organizational structure, potentially enhancing collaboration and retention. Disparities between \emph{Contribution Share (\%)} and \emph{Contributor Share (\%)} in certain countries highlight the impact of individual contributors. For example, in the Rust Analyzer project, Germany's significant contribution share (46.03\%) coupled with a lower contributor share (8.53\%) indicates that a few individuals are responsible for a large portion of contributions. Similarly, Taiwan in the Cargo project shows a high contribution share (29.66\%) from a small contributor base (1.46\%). These findings suggest that active contributions do not fully align with the broader community's demographic distribution and also follow the Pareto principle seen in the Rust ecosystem \cite{free-paid-devs-rust-ecosystem}. 

The concentration of contributions among specific countries and individuals, along with low gender diversity, points to potential barriers to participation for certain groups and regions.

\subsection{Collaboration Patterns and Interaction Dynamics}
\textbf{Autonomous Non-member  Contributor Clusters}: The presence of clusters without any member nodes across all three projects highlights the significant role of non-member contributors in the Rust ecosystem. These independent clusters suggest that non-members can initiate and sustain collaborative efforts without direct involvement from official Rust organization members. This phenomenon underscores the openness of the Rust community and its reliance on contributions from the broader developer population. 

\textbf{Varying Contributor Dynamics}: The central clustering of member nodes, particularly in the Rust and Rust Analyzer projects, indicates that members tend to collaborate closely with one another and may serve as hubs within the network. Their central positions suggest they are integral to the project's development, possibly taking on code review, mentoring, and coordination roles. The dispersion of member nodes in the Cargo project may reflect different collaboration practices or a more decentralized organizational structure. Consequently, the high reciprocity between all the nodes in the Cargo project compared to others is to be noticed. A reciprocity score of 0.5 in the Cargo project implies a higher degree of mutual collaboration, where contributors frequently engage in bidirectional interactions. This indicates equal collaboration dynamics, where contributors both author and review PRs reciprocally. In contrast, the lower reciprocity scores in the Rust and Rust Analyzer projects suggest a more unidirectional collaboration pattern. This may be due to hierarchical structures, where specific contributors predominantly take on reviewing roles, possibly due to their expertise or organizational positions, as observed in the network figures of the respective repositories with dense member hubs. This gives us an idea of how decentralized structures lead to more collaboration, measured through reciprocity in this context. 

However, we should consider collaboration change concerning project size and domain~\cite{empirical-social-teams-github}. The low density observed across all networks aligns with the characteristics of real-world social networks, where individuals tend to have a limited number of connections relative to the total possible. 
Overall, the results suggest that while members play important roles within the Rust projects, non-members are also critical contributors who engage in meaningful collaborations. The varying interaction dynamics across projects reflect differences in domain complexity, collaboration practices, and possibly project governance structures. These insights highlight the need for tailored approaches in community management and contributor support to accommodate the unique characteristics of each project within the Rust ecosystem.

\subsection{Diversity Within Subcommunities}
\textbf{Substantial Geographic Representation:} The high Blau's Index scores in most subcommunities indicate substantial geographical diversity, particularly in the Rust and Rust Analyzer projects. This diversity suggests that contributors from different countries are collaborating effectively within these projects. 

\textbf{Low Gender Representation:} The gender diversity among top contributors is markedly low. The absence of women contributors among the most central and influential nodes across all projects highlights a significant gender disparity. This aligns with broader trends observed in OSS communities, where women's participation is often limited~\cite{bosu2019diversity,trinkenreich2020hidden}. The lack of gender diversity may hinder the inclusivity and representativeness of the projects and points to a need for initiatives to encourage and support women's contributions.

\textbf{Moderate Sustainability:} The presence of members and non-members among the top contributors demonstrates the open and collaborative nature of the Rust ecosystem. Non-members can achieve significant influence and centrality within the network, indicating that official membership is not a prerequisite for impactful contributions. This openness can foster a more diverse and dynamic contributor base, as it lowers barriers to participation. The differences in betweenness centrality scores across the projects highlight variations in network structure and collaboration dynamics. The Rust project's lower betweenness centrality suggests a more distributed network, spreading influence across multiple contributors and reducing dependency on specific individuals. In contrast, the higher betweenness centrality in the Rust Analyzer and Cargo projects indicates that certain contributors act as key connectors or gatekeepers/blockers. This centralization, although showing a group of vital contributors pivotal for the ecosystem, may pose risks if these individuals reduce their participation, potentially impacting the project's progress and collaboration flow. The geographic diversity among top contributors, especially in the Rust Analyzer project, reflects the project's ability to attract and integrate contributors from a variety of countries, including those less represented in the Rust project. Contributors from Romania, Japan, and Iran play significant roles, suggesting that the project has effective mechanisms for engaging and supporting international contributors. Overall, while the Rust ecosystem shows substantial geographical diversity, particularly within subcommunities, there is a clear need to improve gender diversity among contributors. Addressing gender disparities is likely to enhance inclusivity, impact, and innovation. 

With the growth the Rust community faces now, developers are much more likely to be prone to burnout~\cite{burnout-oss-labor}. High centralization in certain projects, as seen in Rust Analyzer and Cargo, may exacerbate this risk by placing undue pressure on key contributors who serve as critical connectors within the network. There is a pressing need~\cite{schueller2022modelinginterconnectedsocialtechnical} to attract and retain new contributors from emerging OSS communities, ensuring a sustainable influx of diverse talent that can drive future innovation~\cite{github-survey-2023}. 


{\setlength{\parskip}{0pt}%
 \titlespacing*{\section}{0pt}{8pt}{6pt}}
 \section{Limitations and Future Work}
 \label{limitations-future-work}
\subsection{Limitations}

While our study provides valuable insights into collaboration and diversity in the Rust community, several limitations must be acknowledged.  
First, the GitHub API imposes limits on the granularity of extractable data, limiting access to detailed user profiles and complete historical records beyond the chosen one-year window. Additionally, strict API rate limits significantly delayed data collection, requiring multiple iterations to capture all relevant pull request information.

Location information was based on self-reported GitHub profile entries and likely reflects contributors' current locations rather than their nationalities, which may limit the accuracy of capturing broader ethnic or racial diversity. Automated tools such as NamSor, while commonly used in prior studies~\cite{qiu2023gender,free-paid-devs-rust-ecosystem}, may also misclassify gender, particularly for gender-neutral or culturally diverse names. Similarly, the geolocation parsing through NLTK introduced reliability concerns, and manual corrections during the verification process carried a risk of human error.

Our mining and analysis scripts, developed in Python and R, could be subject to undetected coding errors, despite careful validation. Furthermore, we excluded self-loops in the constructed collaboration graphs to better focus on contributor interactions, which may slightly influence network density metrics. All network and diversity metrics were derived from a single year of data, offering a contemporary but time-limited view of collaboration patterns.

Finally, the findings are specific to the Rust ecosystem and should be generalized to other OSS communities with caution. The exclusive reliance on quantitative pull request data, without complementary qualitative insights such as interviews or surveys, may leave out important contextual factors influencing collaboration dynamics.

\subsection{Future Work}

Several avenues could be pursued to address these limitations and deepen our understanding of collaboration and diversity in OSS communities. Conducting \textbf{longitudinal studies} across multiple years~\cite{data-rust-ecosystem} would help reveal trends in contributor retention, shifts in engagement, and the effects of organizational or ecosystem changes over time. Building on the insights from this study, organizations like the Rust Foundation\footnote{\url{https://foundation.rust-lang.org/}} could implement targeted initiatives to improve sustainability and inclusivity, such as lowering entry barriers, creating mentorship programs, and providing support for underrepresented groups. These efforts are especially important as many tech firms scale back diversity initiatives.\footnote{https://www.bbc.com/news/articles/cgmy7xpw3pyo}
Expanding this research to include \textbf{multiple OSS communities} across different domains and maturity levels would allow comparative analyses, highlighting common collaboration patterns or unique ecosystem-specific challenges. Additionally, integrating \textbf{further network metrics and analytical frameworks}, such as analyzing motifs, clustering coefficients, or applying machine learning techniques, could offer a deeper structural understanding of collaboration networks.  
Finally, complementing quantitative analyses with \textbf{qualitative research}, through surveys or interviews with contributors and maintainers, could provide critical context behind observed patterns, offering a more comprehensive view of the social and organizational factors shaping OSS collaboration.

\section{Conclusion}
\label{conclusion}

This study investigated the demographic diversity and collaboration dynamics within the Rust programming language community by analyzing PR data from three pivotal repositories: Rust (the core compiler), Rust Analyzer (an IDE front-end), and Cargo (the package manager). Our analysis revealed that active contributors to these projects exhibit lower geographic and gender diversity compared to the broader Rust user community, as indicated by the 2023 Rust Annual Survey. 

Specifically, most of the top contributors are men, and contributions are predominantly concentrated in a few countries. This concentration highlights a disparity between the global user base and the active contributor community, suggesting potential barriers to broader participation. The social network analysis highlighted the importance of non-members contributing substantially despite not being official Rust GitHub members. However, the reliance on a small number of central individuals, particularly in Rust Analyzer and Cargo, poses potential risks to project resilience. 
The diversity within subcommunities demonstrated substantial geographic diversity in most clusters, particularly within the Rust and Rust Analyzer projects. Nonetheless, gender diversity remains significantly low, with very few women contributors among the most central and influential nodes. 

In conclusion, while the Rust ecosystem benefits from a globally active and open contributor base, there are clear disparities in demographic diversity that need to be addressed to foster a more inclusive and sustainable community. 

\section{Data availability}
\label{data-availability}
We have made our anonymized data and scripts publicly available~\cite{reproducibility} to ensure reproducibility and facilitate replication of our findings.

\begin{acks}
This research was supported by the Natural Sciences and Engineering Research Council of Canada (NSERC), Grant RGPIN-2022-03265.
\end{acks}

\bibliographystyle{ACM-Reference-Format}
\bibliography{references}

\end{document}